\documentclass[]{aa}
\def\apj{ApJ}
\def\apjs{ApJS}
\def\aap{A\&A}

\def\mnras{MNRAS}

\def\ueber#1#2{{\setbox0=\hbox{$#1$}%
  \setbox1=\hbox to\wd0{\hss$ #2$\hss}%
  \offinterlineskip
  \vbox{\box1\box0}}{}}

\def\lesssim{\,\lower 1mm \hbox{\ueber{\sim}{<}}\,}
\def\grsim{\,\lower 1mm \hbox{\ueber{\sim}{>}}\,}

\makeatletter

\let\@internalcite\cite
\def\cite{\@ifstar{\citeyear}{\citefull}}
\def\citefull{\def\astroncite##1##2{##1 ##2}\@internalcite}
\def\citeyear{\def\astroncite##1##2{##2}\@internalcite}

\def\citeau{\def\astroncite##1##2{##1}\@internalcite}
\def\citen{\def\astroncite##1##2{##1 (##2)}\@internalcite}
\def\possesivcite{\def\astroncite##1##2{##1's (##2)}\@internalcite}

\def\@citex[#1]#2{\if@filesw\immediate\write\@auxout{\string\citation{#2}}\fi
  \def\@citea{}\@cite{\@for\@citeb:=#2\do
    {\@citea\def\@citea{; }\@ifundefined
       {b@\@citeb}{{\bf ?}\@warning
       {Citation `\@citeb' on page \thepage \space undefined}}%
{\csname b@\@citeb\endcsname}}}{#1}}
\def\@cite#1#2{#1\if@tempswa , #2\fi}
\def\@biblabel#1{}
\makeatother

\title{Monte Carlo simulations of the halo white dwarf population}

\author{Enrique Garc\'{\i}a--Berro\inst{1,2} \and
        Santiago Torres\inst{1} \and
        Jordi Isern\inst{2,3} \and
        Andreas Burkert\inst{4}}
\titlerunning{The halo white dwarf population}
\authorrunning{E. Garc\'\i a--Berro, S. Torres \& J. Isern}

\institute{$^1$Departament de F\'\i sica Aplicada, Escola  Polit\'ecnica 
               Superior de Castelldefels, Universitat   Polit\`ecnica de 
               Catalunya,   Avda.  del  Canal  Ol\'\i  mpic  s/n,  08860 
               Castelldefels, Spain\\
           $^2$Institute for Space Studies of Catalonia, c/Gran Capit\`a
               2--4, Edif.  Nexus 104, 08034 Barcelona, Spain\\
           $^3$Institut de Ci\`encies de l'Espai, C.S.I.C.\\
           $^4$Max-Planck-Institut   f\"ur  Astronomie,  Koenigstuhl 17, 
               69117 Heidelberg, Germany}

\offprints{E. Garc\'\i a--Berro}
\date{\today}

\abstract{The interpretation of microlensing results towards the Large
Magellanic  Cloud (LMC)  still remains  controversial.   Whereas white
dwarfs  have been  proposed to  explain these  results and,  hence, to
contribute significantly to  the mass budget of our  Galaxy, there are
as well  several constraints on the  role played by  white dwarfs.  In
this  paper  we  analyze  self-consistently  and  simultaneously  four
different  results,  namely, the  local  halo  white dwarf  luminosity
function, the microlensing results  reported by the MACHO team towards
the LMC, the results of Hubble Deep Field (HDF) and the results of the
EROS experiment, for several initial  mass functions and halo ages. We
find  that  the proposed  log-normal  initial  mass  functions do  not
contribute  to solve the  problem posed  by the  observed microlensing
events and,  moreover, they overproduce white dwarfs  when compared to
the results of the  HDF and of the EROS survey. We  also find that the
contribution of  hydrogen-rich white dwarfs  to the dynamical  mass of
the halo of the Galaxy cannot be more than $\sim 4\%$.
 }

\begin{document}

\maketitle


\section{Introduction}

White  dwarfs  are the  most  common  remnants  of stellar  evolution.
Moreover, since  white dwarfs are long-lived objects  and the physical
processes governing their evolution are relatively well understood ---
at least up to moderately low luminosities --- they provide us with an
invaluable tool  to study the  evolution and structure of  our Galaxy.
In  fact, the  disk  white  dwarf luminosity  function  has become  an
important tool to determine some properties of the local neighborhood,
such as its  age (Winget et al.  1987; Garc\'\i  a-Berro et al.  1988;
Hernanz et al.  1994), or the  past history of the star formation rate
(Noh  \&  Scalo  1990; D\'\i  az-Pinto  et  al.   1994; Isern  et  al.
1995a,b).   This  has  been  possible  because now  we  have  improved
observational  luminosity  functions  (Liebert,  Dahn \&  Monet  1988;
Oswalt et  al.  1996; Leggett, Ruiz  \& Bergeron 1998)  and because we
have reliable cooling sequences ---  see, for instance, Salaris et al.
(2000), and references therein.

\begin{table*}
\centering
\begin{tabular}{ccl}
\hline
\hline
{\rm IMF} & $\phi(M)$ & {\rm Reference} \\
\hline
 & & \\
{\rm Standard} & $
\left\{\begin{array}{lcl}
(M/M_{\odot})^{-1.2} & {\rm for }& M<1.0\,M_{\odot} \\
 & & \\
(M/M_{\odot})^{-2.7} & {\rm for }& 1.0\,M_{\odot}<M<10.0\,M_{\odot} \\
 & & \\
0.4(M/M_{\odot})^{-2.3} & {\rm for }& M>10.0\,M_{\odot} \\
\end{array}
\right. $
& {\rm Scalo (1998)} \\
 & & \\
{\rm CSM1} & 
$\left\{
\begin{array}{c} 
\exp\left[-(M_{\rm cut}/M)^{\beta_1}(M/M_{\odot})^{\beta_2}\right] \\
M_{\rm cut}=2.0\, M_{\odot}$, $\beta_1=2.2 {\rm \ y\ }  
\beta_2=5.15 
\end{array}
\right.
$  & Chabrier et al. (1996) \\
 & & \\
{\rm CSM2} & 
$
\left\{
\begin{array}{c} 
\exp\left[-(M_{\rm cut}/M)^{\beta_1}(M/M_{\odot})^{\beta_2}\right] \\
M_{\rm cut}=2.7\, M_{\odot}$, $\beta_1=2.2 {\rm \ y\ }  
\beta_2=5.75 
\end{array}
\right.
$  & Chabrier et al. (1996) \\
 & & \\
{\rm AL} & 
$
\left\{
\begin{array}{c} 
\exp\left[1-\frac{1}{2\sigma_2}(\log(\frac{M}M_{\rm cut})^{2}\right] \\
M_{\rm cut}=2.3\, , {\rm \ y\ }  
\sigma=0.44
\end{array}
\right.
$  & Adams \& Laughlin (1996) \\
 & & \\
\hline
\hline
\end{tabular}
\caption{The different IMFs used in this paper.}
\end{table*}

Although the situation for the disk white dwarf population seems to be
already clear and  well understood, this is not the  case for the halo
white  dwarf  population.   Moreover,  the  discovery  of  microlenses
towards the Large Magellanic Cloud  (Alcock et al. 1996, Alcock et al.
2000; Lasserre et  al.  2001) has generated a  large controversy about
the  possibility that  white  dwarfs could  be  responsible for  these
microlensing   events   and,  thus,   could   provide  a   significant
contribution to  the mass budget of our  Galactic halo.  Nevertheless,
white dwarfs as dark matter candidates are not free of problems, since
an  excess  of  this  kind  of  objects  would  necessarily  imply  an
overproduction of  low--mass main  sequence red dwarfs  and high--mass
stars that could  eventually explode as Type II  supernovae.  In order
to  solve  these  problems  Adams  \& Laughlin  (1996)  and  Chabrier,
S\'egretain  \&  M\'era  (1996)  proposed non--standard  initial  mass
functions in which  the formation of both low and  high mass stars was
suppressed.   Besides  the  lack  of evidences  favoring  such  ad-hoc
initial  mass functions,  they are  not free  of problems  either.  In
particular, the  formation of  a typical ($M  \sim \,  0.6\, M_\odot$)
white  dwarf is  accompanied by  the injection  into  the interstellar
medium  of  a  sizeable amount  of  mass  ($\sim  1.5 \,  M_\odot$  on
average).  Since  Type II supernovae are suppressed  in ad-hoc initial
mass functions, there  is not enough energy to  eject this matter into
the intergalactic medium and  the ejected mass, which is significantly
contaminated  by metals  (Abia et  al.  2001;  Gibson \&  Mould 1997),
cannot be accomodated into the  Galaxy (Isern et al.  1998).  Finally,
an excess of  white dwarfs also translates into  an excess of binaries
containing  such  stars  with  the  subsequent  increase  of  Type  Ia
supernova  rates which,  ultimately,  results in  an  increase in  the
abundances  of  the  elements  of  the  iron  peak  (Canal,  Isern  \&
Ruiz--Lapuente 1997).  All these  arguments have forced the search for
other  possible explanations,  such as  self--lensing in  the  LMC (Wu
1994; Salati et  al.  1999), or background objects  (Green \& Jedamzik
2002) which have not been yet totally discarded.

The  suggestion of the  MACHO team  (Alcock et  al.  1997,  2000) that
white  dwarfs  contribute significantly  to  the  mass  budget of  the
Galactic halo  has motivated a large number  of observational searches
(Knox, Hawkins \& Hambly 1999;  Ibata et al.  1999; Oppenheimer et al.
2001; Majewski \& Siegel 2002,  Nelson et al.  2002) for these elusive
white  dwarfs.   Also several  theoretical  works  (Reyl\'e, Robin  \&
Crez\'e 2001;  Flynn, Holopainen \& Holmberg 2003)  have analyzed this
possibility.  However, the controversy  of whether or not white dwarfs
can provide a significant contribution  to the Galactic dark matter is
still open and  deserves some more attention.  In  Isern et al. (1998)
we  analyzed  the halo  white  dwarf  population.   In particular,  we
computed, assuming a standard initial mass function and updated models
of  white dwarf  cooling, the  expected luminosity  function,  both in
luminosity  and  in visual  magnitude,  for  different star  formation
rates. We  showed that a  deep enough survey (limiting  magnitude $\ga
20$) could  provide important information  about the halo age  and the
duration of  the formation  stage. We also  showed that the  number of
white  dwarfs  produced  using  the  proposed biased  IMFs  could  not
represent  a large  fraction  of the  halo  dark matter  if they  were
constrained by the observed  luminosity function of halo white dwarfs.
However, within  the approach adopted  there the biases  introduced by
the  sample selection  procedures were  not taken  into  account. More
recently,  we  have analyzed  (Torres  et  al.   2002) the  sample  of
Oppenheimer et  al.  (2001).  In this  paper we examine  in detail the
results of the MACHO team  (Alcock et al.  2000) carefully taking into
account the observational biases, thus updating our previous papers on
this subject.  We also study the number of white dwarfs which could be
potentially found  in the HDF (Ibata  et al.  1999).   Finally we also
analyze  the  very  recent  observational  results of  the  EROS  team
(Goldman et al.  2002), which set  a very stringent upper limit to the
white dwarf content of the Galactic halo.  All these analyses are done
by making  use of  a Monte Carlo  simulator (Garc\'\i a--Berro  et al.
1999; Torres  et al.  1998).  This  is an important  issue since white
dwarf  populations  are  usually  drawn  from  kinematically  selected
samples   (white  dwarfs   with  relatively   high   proper  motions).
Therefore,  some kinematical  biases or  distortions are  expected.  A
Monte  Carlo simulation  of  a  model population  of  white dwarfs  is
expected to  allow the  biases and effects  of sample selection  to be
taken  into account, so  the properties  of the  real sample  could be
corrected --- or, at least,  correctly interpreted --- provided that a
detailed simulation from the very  early stages of source selection is
performed accurately.  Our paper  is organized as follows.  In section
\S 2  we describe in  full detail  the Monte Carlo  code.  In \S  3 we
present the results, whereas in \S 4 our conclusions are summarized.


\section{The model}

In this  section we  discuss the main  ingredients of our  Monte Carlo
simulator.  Since we want to self-consistently simulate simultaneously
four different results, namely,  the local halo white dwarf luminosity
function, the microlensing results towards the LMC, the results of the
HDF,  and  the  results of  the  EROS  team,  and  each one  of  these
simulations requires slightly different  inputs, we will describe them
in  separate subsections.   All of  them, however,  share  some common
ingredients like  a random  number generator, which  is always  at the
heart of  any Monte  Carlo simulation.  We  have used a  random number
generator algorithm (James 1990)  which provides a uniform probability
density within the interval $(0,1)$ and ensures a repetition period of
$\grsim 10^{18}$, which  is enough for our purposes.   Each one of the
Monte Carlo  simulations discussed in  section 3 below consists  of an
ensemble of  40 independent realizations of the  synthetic white dwarf
population, for which the  average of any observational quantity along
with  its corresponding  standard deviation  were computed.   Here the
standard deviation  means the ensemble mean of  the sample dispersions
for a typical sample.

\subsection{The local halo white dwarf luminosity function}

We have considered  a typical spherically symmetric halo.   One of the
most commonly used models of  this type is the isothermal sphere.  The
density profile of the luminous halo is given by the law

\begin{equation}
\rho(r)=\rho_0\frac{a^2+R_{\odot}^2}{a^2+r^2}
\end{equation}

\noindent where $a\approx  5$ kpc is the core  radius, $\rho_0$ is the
local  halo  density and  $R_{\odot}=$8.5  kpc  is the  galactocentric
distance  of  the Sun.   We  randomly  choose  three numbers  for  the
spherical  coordinates $(r,\theta,\phi)$  of each  star of  the sample
within  approximately 350  pc from  the  sun, according  to Eq.   (1).
Afterwards we draw another pseudo-random number in order to obtain the
main sequence  mass of each star,  according to one of  the four model
initial mass functions that will  be studied here.  These initial mass
functions are,  respectively, a standard IMF (Scalo  1998), the biased
IMF of  Adams \& Laughlin (1996)  and the two ad-hoc  IMFs proposed by
Chabrier et al.   (1996).  All these mass functions  are summarized in
table 1.  Once the mass of  the progenitor of the white dwarf is known
we randomly choose the time at which each star was born ($t_{\rm b}$).
For this purpose  we assume that the halo was formed  14 Gyr ago (see,
however, section \S 3 below) in  an intense burst of star formation of
duration $\sim 1$~Gyr.  Given the age of the halo, $t_{\rm b}$ and the
main sequence lifetime as a function  of the mass in the main sequence
(Iben \&  Laughlin 1989) we know  which stars have had  time enough to
become white dwarfs, and given  a set of cooling sequences (Salaris et
al.   2000)  and the  initial  to  final  mass relationship  (Iben  \&
Laughlin 1989), which are their luminosities.

The velocity distribution has been modeled according to a Gaussian law
(Binney \& Tremaine 1987):

\begin{equation}
f(v_r,v_t)=\frac{1}{(2\pi)^{3/2}}\frac{1}{\sigma_r\sigma_t^2}
\exp\left[-\frac{1}{2}\left(\frac{v_r^2}{\sigma_t^2}+\frac{v_{t}^2}
{\sigma_t^2}\right)\right]
\end{equation}

\noindent  where $\sigma_r$  and  $\sigma_t$ ---  the  radial and  the
tangential velocity  dispersion, respectively  --- are related  by the
following expression:

\begin{equation}                                 
\sigma_t^2=\frac{V_{\rm c}^2}{2}+\left[1-\frac{r^2}{a^2+r^2}\right]
\sigma_r^2+\frac{r}{2}\frac{{\rm d}(\sigma_r^2)}{{\rm d}r}
\end{equation}

\noindent  which,  in  a  first  approximation,  leads  to  $\sigma_r=
\sigma_t= {V_{\rm  c}}/{\sqrt{2}}$.  --- see, for  instance, Binney \&
Tremaine (1987).  For the calculations reported here we have adopted a
circular  velocity $V_{\rm  c}= 220$~km/s.   From these  velocities we
obtain the heliocentric  velocities by adding the velocity  of the LSR
$v_{\rm  LSR}=-220  $~km/s  and  the  peculiar velocity  of  the  sun:
$(U_\odot,  V_\odot,W_\odot)=(10.0,15.0,8.0) $~km/s (Dehnen  \& Binney
1998).  Since white  dwarfs usually do not have  determinations of the
radial component  of the velocity,  when needed for  the observational
comparison  the  radial velocity  is  eliminated.   Moreover, we  only
consider stars with tangential velocities  in the range $ 250 \lesssim
v_{t} \lesssim 750$~km/s.  Stars with velocities smaller than 250~km/s
would not be considered as halo members, whereas stars with velocities
larger  than 750~km/s would  have velocities  exceeding 1.5  times the
escape velocity, which we obtain from Binney \& Tremaine (1987):

\begin{equation}
v_{\rm e}^2=2V_c^2[1+\ln(r_*/r)]    
\end{equation}

\noindent where $r_*\simeq 41$ kpc is the radius of the galactic halo.

In  order to  build  the  white dwarf  luminosity  function using  the
$1/V_{\rm max}$ method (Schmidt 1968) a smaller sample of white dwarfs
must be culled from the original sample  and in order to do this a set
of restrictions in visual magnitude and proper motion must be adopted.
The  restriction  in  magnitude  will  be discussed  in  \S  3  below.
Regarding   the  proper   motion   cut  we   have   chosen  $\mu   \ge
0.16^{\prime\prime}$~yr$^{-1}$ as it was done in Oswalt et al.  (1996)
and in Garc\'\i a--Berro et al.  (1999).  Besides, the $1/V_{\rm max}$
method  requires that  all  the objects  belonging  to the  restricted
sample must have known parallaxes.   This, in turn, means that all the
white dwarfs belonging to this sample are within a sphere of radius of
roughly 200 pc centered in the location of the sun.

\begin{figure*}
\vspace{11cm}
\includegraphics{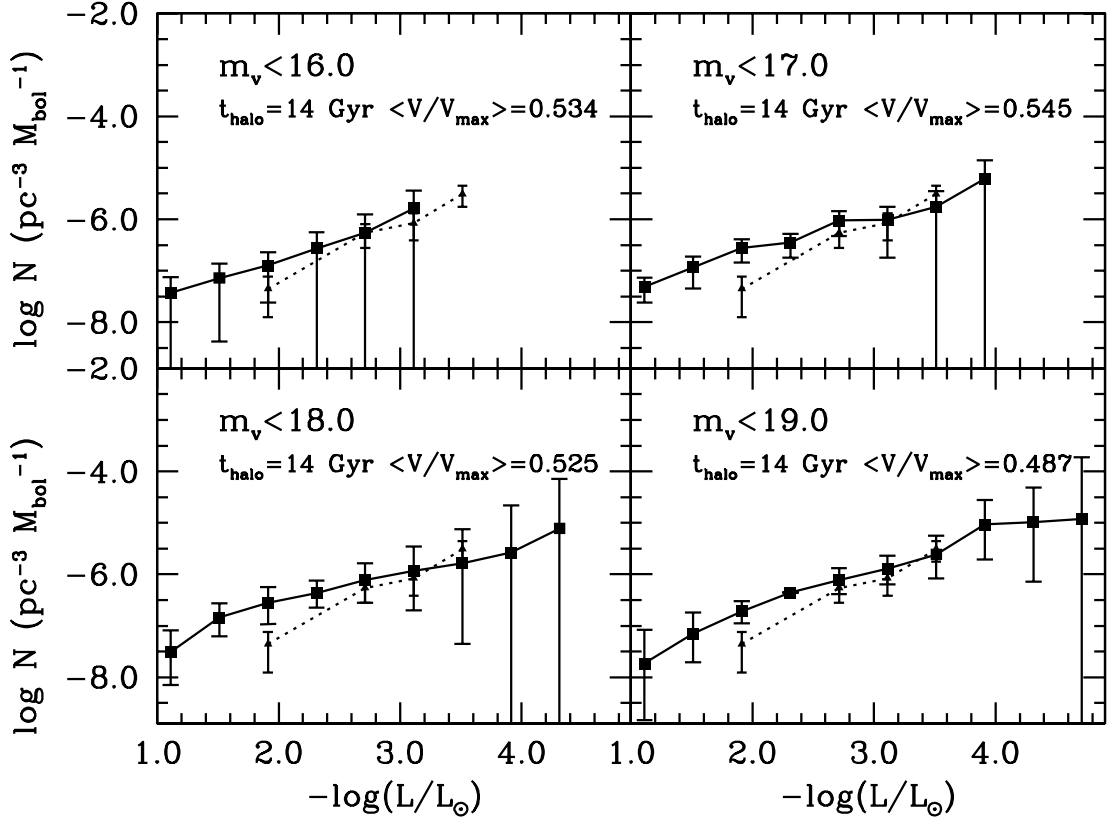}
\caption{Luminosity function of halo white dwarfs for several limiting
	magnitudes.}
\end{figure*}

\subsection{Microlensing events towards the LMC}

In order  to produce a set  of microlensing events towards  the LMC we
also  need  to  simulate   the  characteristics  of  the  white  dwarf
population towards  the LMC.  For  this purpose we generate  the three
galactic  coordinates $(r,l,b)$ of  the white  dwarfs of  the galactic
halo inside a small pencil of $4^{\circ} \times 4^{\circ}$ centered in
the LMC location, $(l,b)= (280^{\circ},-33^{\circ})$.  The $l$ and $b$
distributions  are  practically uniform  in  this  small window.   The
radial coordinate is  always smaller than the outer  limit of the halo
($r<41$~kpc) and according to the radial distribution

\begin{equation}
\rho=\rho_0(R_\odot/r)^\gamma
\end{equation}

\noindent where $\gamma=3.4$ --- see, however, \S 3.7.  We have chosen
this distribution instead of that  of equation (1) because in this way
the  number of  microlensing events  is maximized.   In fact  the mass
distribution  of  microlenses does  not  necessarily  follow the  mass
distribution of the  luminous halo.  However, in a  first step (see \S
3.6 below),  we have normalized  the density of white  dwarfs obtained
from  this  distribution to  the  white  dwarf  density of  the  local
neighborhood,   $n\sim  9.0\times   10^{-6}$~pc$^{-3}$   for  $\log(L/
L_\odot)\ga -3.5$ (Torres et  al.  1998).  The velocity dispersions in
this  case are  determined from  Markovi\'c \&  Sommer--Larsen (1996).
For the radial velocity dispersion we have:

\begin{equation}
\sigma_r^2=\sigma_0^2+\sigma_+^2\left[\frac{1}{2}
-\frac{1}{\pi}\arctan\left(\frac{r-r_0}{l}\right)\right]
\end{equation}

\noindent where $\sigma_0=80\, {\rm km\,s^{-1}}$, $\sigma_+=145\, {\rm
km\,s^{-1}}$,  $r_0=10.5$   kpc  and  $l=5.5$   kpc.   The  tangential
dispersion is given by:

\begin{equation}
\sigma_t^2=\frac{1}{2}V_{\rm c}^2-\left(\frac{\gamma}{2}-1\right)\sigma_r^2+
\frac{r}{2}\frac{{\rm d}\sigma_r^2}{{\rm d}r}
\end{equation}

\noindent where
\begin{equation}
r\frac{{\rm d}\sigma_r^2}{{\rm d}r}=
-\frac{1}{\pi}\frac{r}{l}\frac{\sigma_+^2}{1+[(r-r_0)/l]^2}
\end{equation}

In  order to  decide which  white  dwarf could  potentially produce  a
microlensing  event a  magnitude cut  must be  adopted.  If  the white
dwarf is brighter than this limiting magnitude it could be potentially
detected and, consequently, would not be a genuine microlensing event.
We have explored a wide range of possibilities, namely $m_{\rm V}^{\rm
cut}= 17.5^{\rm mag}, \, 22.5^{\rm mag}$ and $27.5^{\rm mag}$, but, as
a fiducial value, we have taken $m_{\rm V}^{\rm cut}= 17.5^{\rm mag}$,
which is the value adopted by  Alcock et al.  (2000).  We also need to
simulate  the population  of stars  of the  LMC itself.   In  order to
produce such a population we  distribute the number of monitored point
sources ($\sim 1.2 \times 10^7$) according to the LMC model of Gyuk et
al.  (2000). That is, we have adopted a scale lenght, $R_{\rm d}=1.57\
{\rm kpc}$, a scale height  $z_{\rm d}=0.3\ {\rm kpc}$, an inclination
angle $i=30^{\circ}$, a distance  $L=50\ {\rm kpc}$, a position angle,
$\phi=170^{\circ}$,  and a  tangential heliocentric  velocity, $v_{\rm
t}=336\  {\rm km s^{-1}}$.   The next  step is  to determine  if there
exists a white dwarf in the line of sight to a given simulated star in
the  LMC.   We  consider  a   white  dwarf  to  be  responsible  of  a
microlensing event if the angular distance between the white dwarf and
the monitored  star is smaller  than the Einstein  radius $\theta_{\rm
E}=R_{\rm E}/D_{\rm OL}$, where $R_{\rm E}$ is the Einstein radius and
$D_{\rm OL}$ is the distance between the observer and the lens.  This,
of course, can  happen at any time during  the total monitoring period
of 5.7 yr,  due to the proper motion of the  lenses, which is obtained
from the previous equations.

The optical depth is obtained,  following Alcock et al.  (2000), using
the expression

\begin{equation}
\tau=\frac{1}{E}\frac{\pi}{4}\sum_i \frac{\hat t_i}{\varepsilon(\hat t_i)}
\end{equation}

\noindent  where  $E=6.12\times  10^7$   is  the  total  exposure  (in
star-years), $\hat  t_i$ is the Einstein ring  diameter crossing time,
and  $\varepsilon(\hat   t_i)$  is  the   detection  efficiency.   The
detection efficiency has been modelled as:

\begin{equation}
\varepsilon(\hat{t})=
\left\{
\begin{array}{cc}
0.43\,{\rm e}^{-(\ln(\hat{t}/T_{\rm m}))^{3.58}/ 0.87}, & \hat{t}>T_{\rm m} \\
0.43\,{\rm e}^{-|\ln(\hat{t}/T_{\rm m})|^{2.34}/11.16}, & \hat{t}<T_{\rm m} 
\end{array}
\right.
\end{equation}

\begin{figure*}
\vspace{11cm}
\includegraphics{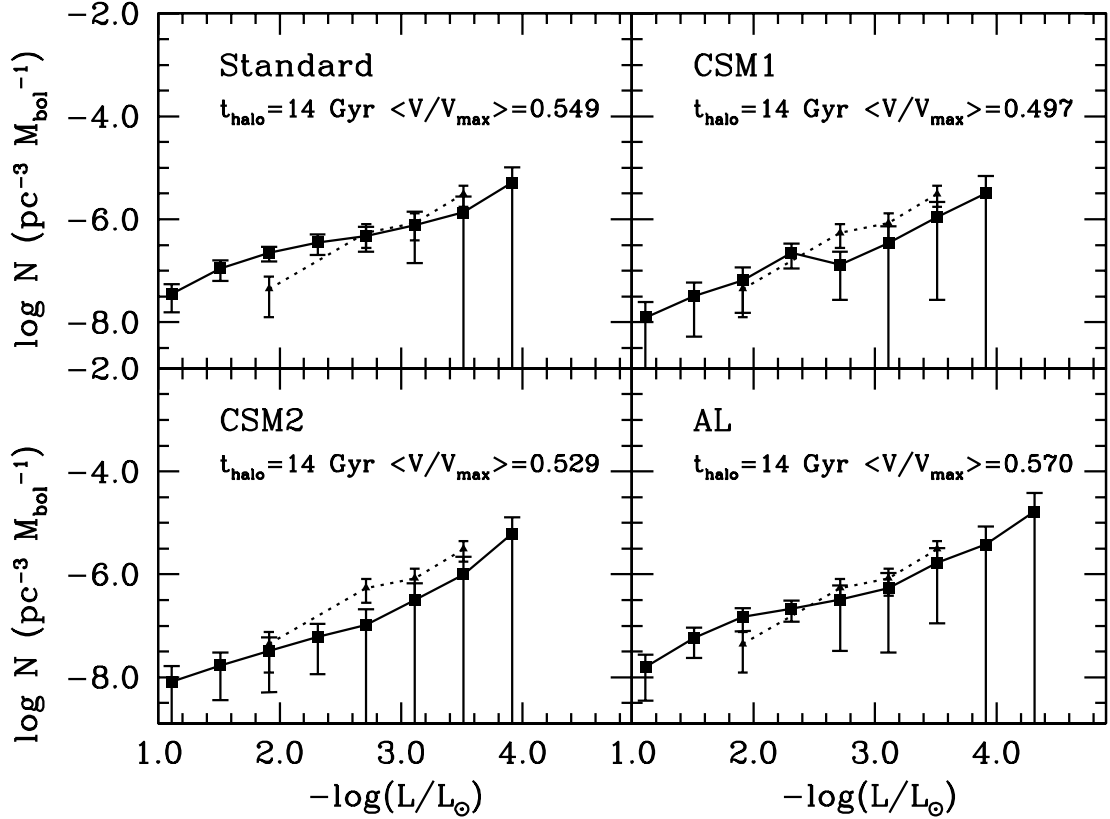}
\caption{Luminosity function of halo white dwarfs for several IMFs.}
\end{figure*}

\noindent where $T_{\rm m}=250$ days.  This expression provides a good
fit to the results of Alcock et al.  (2000).

\subsection{The Hubble Deep Field simulation}

For the simulation of the HDF we have distributed stars in a window of
$1.2^{\circ}  \times 1.2^{\circ}$  centered  around $(l,b)=(125^\circ,
55^\circ)$.  The radial distribution  is, again, according to equation
(5) within  the outer  halo  limit.  The  velocities are  consequently
drawn from  equations (2)  and (6) to  (8).  This simulation  has been
also normalized to the local density given by the halo simulation.

\begin{figure*}
\vspace{11cm}
\includegraphics{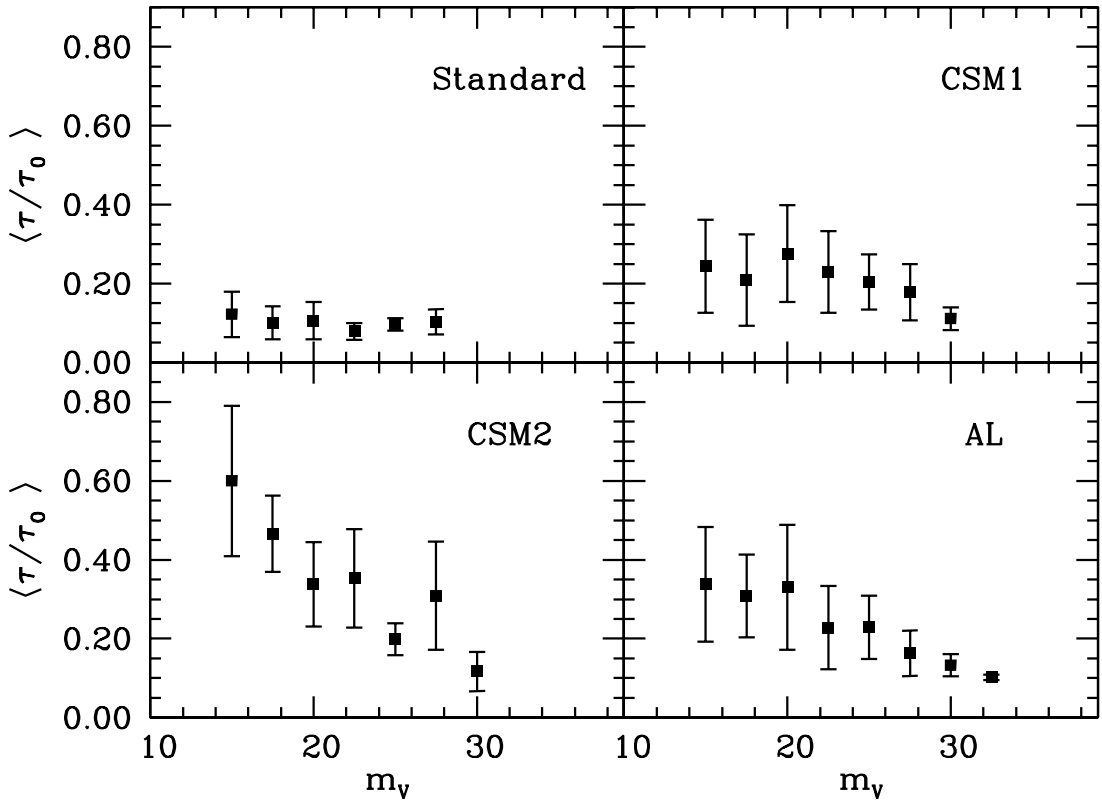}
\caption{Microlensing optical  depth towards the LMC as  a function of
         the limiting magnitude for several IMFs.}
\end{figure*}

\subsection{The EROS simulation}

Finally, for  the simulation of  the EROS results we  have distributed
stars  in  a  window  of  ${190^\circ}^2$  in  the  Southern  Galactic
Hemisphere ($-79^\circ  < b_{\rm  gal} <-48^\circ$), in  the following
strips along the $\alpha$ coordinate, $\Delta \delta =1.4^\circ$ wide:
$22^{\rm h}\,16^{\rm  min}\, <\alpha < \, 3^{\rm  h}\,44^{\rm min}$ at
$\delta=  -44^{\circ}  \,45^\prime$,  $23^{\rm  h}\,31^{\rm  min}\,  <
\alpha  < \,  1\,^{\rm h}\,  34^{\rm min}$  at  $\delta= -40^{\circ}\,
09^\prime$ and $22^{\rm  h} \, 24^{\rm min} \,  <\alpha \leq 3^{\rm h}
\,   28^{\rm   min}$  at   $\delta=   -38^{\circ}\,  45^\prime$,   and
${187^\circ}^2$  in the  Northern Galactic  Hemisphere  ($41^{\circ} <
b_{\rm gal} < 59^{\circ}$) with  $10^{\rm h}\,57^{\rm min} \, < \alpha
\leq 13^{\rm h} \, 23^{\rm min}$ at $\delta= -12^{\circ}$ and $10^{\rm
h}\,  57^{\rm min}  \,  < \alpha  <  12^{\rm h}  \,  53^{\rm min}$  at
$\delta= -4^{\circ}  \, 36^\prime$.   The radial distribution  and the
velocity  distribution   are,  again,  the  same  used   for  the  HDF
simulation.  Also, the density of  white dwarfs has been normalized to
the local density given by the halo simulation.


\section{Results}

\subsection{The halo white dwarf luminosity function}

One of  the most serious problems  that is found  when determining the
observational white dwarf luminosity  function of halo white dwarfs is
that  the real  limiting magnitude  used  in these  studies is  highly
uncertain.  We  have conducted a series of  simulations with different
limiting magnitudes to determine which would be the limiting magnitude
able to reproduce the  observational luminosity function.  In figure 1
we  show the  white dwarf  luminosity functions  obtained  for several
limiting magnitudes  (16, 17, 18  and $19^{\rm mag}$) as  solid lines.
The luminosity function of Torres et  al.  (1998) is shown as a dashed
line for comparison  purposes.  Each panel is clearly  marked with its
corresponding  limiting magnitude.  The  adopted halo  age in  all the
cases was  14 Gyr.  Also  the adopted IMF  is the standard one  in all
four simulations.  The error bars of each luminosity bin were computed
according to Liebert et al.   (1988): the contribution of each star to
the total  error budget in its  bin is conservatively  estimated to be
the same amount that contributes to the resulting density; the partial
contributions of each star in the  bin are squared and then added, the
final error  being the square root  of this value.   This procedure is
followed  for each  one  of the  40  realizations of  the Monte  Carlo
simulation.  After doing this  the ensemble average of the dispersions
is computed.  Obviously the larger the magnitude limit the fainter the
white dwarfs we detect.  If  we disregard as non significative the bin
in which  we only detect  on average one  white dwarf we see  that the
limiting  magnitude that  best reproduces  the luminosity  function of
Torres  et   al.   (1998)  is  $m_{\rm   V}^{\rm  lim}=17^{\rm  mag}$.
Therefore for  the rest  of the simulations  we adopt this  value.  It
should be mentioned as well that in order to detect the cut-off of the
luminosity function  a limiting magnitude of $25^{\rm  mag}$ should be
adopted.  It is worth noting at this point that since the heliocentric
velocity of halo  white dwarfs is considerable, the  proper motion cut
plays a limited role.  In fact  the proper motion cut only affects the
total number  of white dwarfs in the  sample but not the  shape of the
luminosity function.   In contrast, as we  have seen, this  is not the
case for the  cut in magnitude.  This  is the same as to  say that the
proper motion cut  equally affects all the luminosity  bins.  It is as
well interesting to note here that the value of the limiting magnitude
that we have  found so far is in close agreement  with the cut adopted
by Alcock et al.  (2000) for  the monitoring of stars within the MACHO
project, which is $17.5^{\rm mag}$.

\begin{figure*}
\vspace{11cm}
\includegraphics{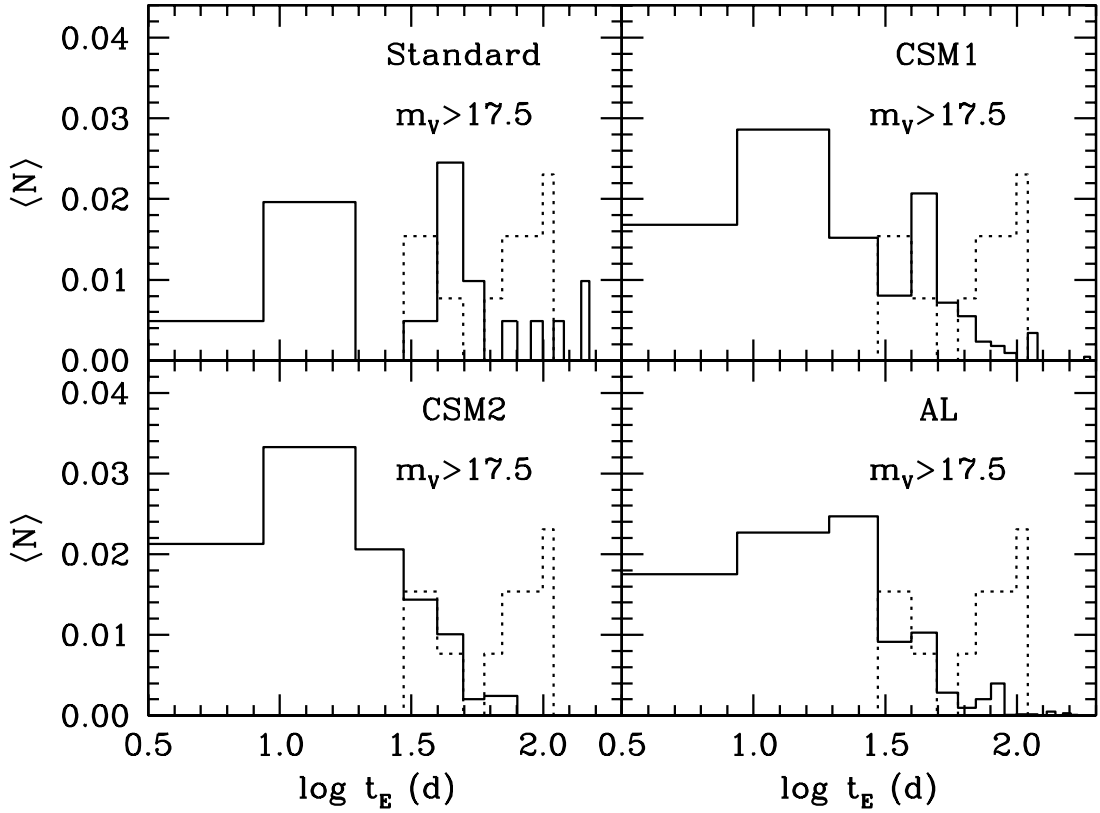}
\caption{Distribution of Einstein  crossing times for the microlensing
	events towards  the LMC of the 40  independent realizations of
	the simulated  population (solid lines) for  several IMFs, and
	of the observational data (dashed lines).}
\end{figure*}

\begin{table*}
\centering
\begin{tabular}{lrrrrrrrrrrrr}
\hline
\hline
\multicolumn{1}{c}{\ } & 
\multicolumn{3}{c}{Standard} &
\multicolumn{3}{c}{CSM1} &
\multicolumn{3}{c}{CSM2} &
\multicolumn{3}{c}{AL} \\
\hline
Magnitude 
& 17.5 &  22.5  & 27.5 
& 17.5 &  22.5  & 27.5  
& 17.5 &  22.5  & 27.5 
& 17.5 &  22.5  & 27.5\\
\cline{2-4} \cline{5-7} \cline{8-10} \cline{11-13}
$\langle N_{\rm WD} \rangle$ & $0\pm 1$ & $0\pm 1$ & $0\pm 1$ & 
$2\pm 4$ & $1\pm 3$ & $1\pm 2$ & $7\pm 3$ & $6\pm 4$ & $2\pm 2$ & 
$5\pm 5$ & $3\pm 3$ & $1\pm 1$\\
$ \langle m \rangle$ $(M/M_{\odot})$ & 0.564 & 0.568 & 0.556 &
0.590 & 0.582 & 0.601 & 0.598 & 0.613 & 0.595 &
0.622 & 0.636 & 0.614  \\
$\langle \mu\rangle$ $(''\,{\rm yr}^{-1})$ &
0.016 & 0.014 & 0.012 & 
0.030 & 0.024 & 0.010  &
0.093 & 0.035 & 0.021 & 
0.035 & 0.030 & 0.012 \\
$\langle d\rangle$ (kpc) &
3.09 & 3.14 & 5.05 & 
1.53 & 2.12 & 6.19  &
0.51 & 1.50 & 3.04  & 
1.63 & 1.65 & 5.10 \\
$\langle V_{\rm tan} \rangle$ $({\rm km\,s}^{-1})$ &
235 & 207 & 285 & 
222 & 243 & 291  &
224 & 251 & 308 &  
272 & 232 & 293  \\
$\langle \hat{t}_{\rm E}\rangle $ (d) & 
61.3 & 75.8& 89.3 & 
36.3 & 52.6 & 70.8  & 
21.7 & 41.7 & 70.9 & 
28.8 & 42.7 & 69.2 \\
$\langle \tau/\tau_0 \rangle$ &
0.101 & 0.079 & 0.102 & 
0.209 & 0.230 & 0.177&
0.466 & 0.353 & 0.308 & 
0.308 & 0.228 & 0.163  \\
\hline
\hline
\end{tabular}
\caption{Summary  of  the  results  obtained  for  the  simulation  of
	microlenses towards the LMC for  an age of the halo of 14~Gyr,
	different model IMFs, and several magnitude cuts.}
\end{table*}

\subsection{The role of the IMF}

In  figure  2  we  show  the  luminosity  functions  obtained  with  a
Salpeter-like  mass  function, the  two  log-normal  IMFs proposed  by
Chabrier et  al.  (1996) --- CSM1  and CSM2, respectively  --- and the
IMF of Adams \& Laughlin (1996) --- AL.  As this figure clearly shows,
the derived luminosity functions are not very sensitive to the precise
shape  of the  IMF.  Moreover,  the completeness,  as measured  by the
$\langle V/V_{\rm max}\rangle$ method, seems to be similar in all four
cases.  Only in the cases CSM1  and, more apparently, in the CSM2 case
there is a slight  underproduction of luminous white dwarfs.  However,
the  significance  is  only  marginal  and,  therefore,  in  order  to
constrain  the IMF  of the  galactic halo  using  intrinsically bright
white dwarfs  deeper surveys  are needed.  We  will come back  to this
issue when studying the HDF simulation in \S 3.4 below.

As previously stated in \S  2.1, all the luminosity functions obtained
here have  been normalized to the  local density of  halo white dwarfs
obtained by Torres et al.  (1998), $n\sim 9.0\times 10^{-6}$~pc$^{-3}$
for $\log(L/ L_\odot)\ga -3.5$, which  for a typical value of the mass
of white dwarfs  ($\simeq 0.6\, M_\odot$) corresponds to  a density of
halo baryonic  matter in the form  of white dwarfs  of $\sim 7.2\times
10^{-6}\,  M_\odot$~pc$^{-3}$.  However, from  our simulations  we can
derive which  would be the {\sl  total} density of  baryonic matter in
the  galactic   halo  within   300  pc  from   the  sun.    We  obtain
$\rho_0=2.2\times 10^{-4}\,  M_\odot$~pc$^{-3}$ for the  standard IMF,
$2.0\times 10^{-3}\, M_\odot$~pc$^{-3}$  for the CSM1 case, $1.1\times
10^{-2}\,  M_\odot$~pc$^{-3}$ for the  CSM2, and  $2.7\times 10^{-3}\,
M_\odot$~pc$^{-3}$  for the  AL  simulation.  These  values, in  turn,
correspond to a fraction $\eta$ of baryonic dark matter of 0.03, 0.25,
1.40  and 0.33,  respectively.   As  it can  be  seen the  differences
between all  the IMFs analyzed  here are considerable.   For instance,
for the CSM2  case we would have more matter  than needed, whereas the
CSM1  and AL mass  functions would  lock a  sizeable fraction  of dark
matter in the form of  main sequence stars, stellar remnants {\sl and}
in  the corresponding  ejected  mass. Finally  the  standard IMF  only
allows for  a modest 3\%  of the required  dark matter. In  this case,
moreover,  roughly 1/3  of the  stellar content  corresponds  to white
dwarfs.

\begin{figure*}
\vspace{11cm}
\includegraphics{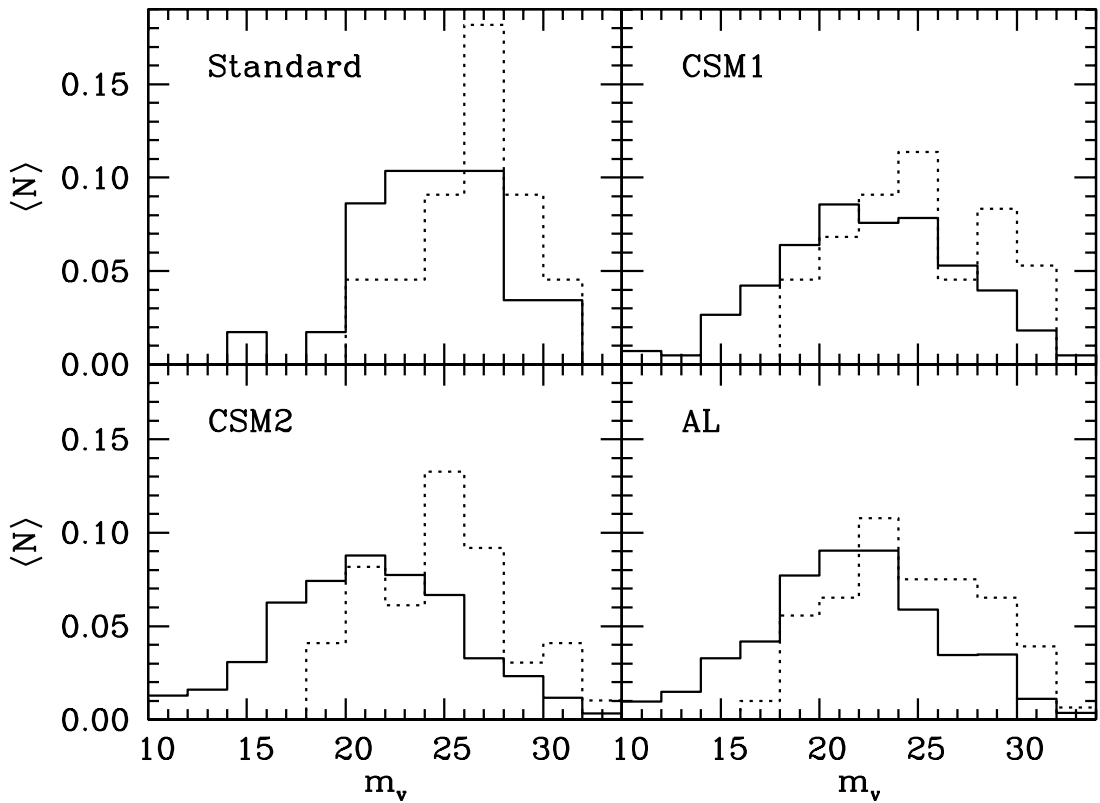}
\caption{Distribution of  the whole  white dwarf population  --- solid
	lines --- and of the white dwarfs responsible for microlensing
	events towards the  LMC --- dashed lines ---  as a function of
	their visual magnitude for several IMFs.}
\end{figure*}

\subsection{Microlenses towards the LMC}

In figure  3 we  show the microlensing  optical depth towards  the LMC
obtained in our simulations normalized  to the value derived by Alcock
et  al.  (2000),  $\tau_0=1.2\times  10^{-7}$, as  a  function of  the
magnitude cut for detection of  white dwarfs (see section \S 3.1).  As
it can be seen in this  figure the contribution of white dwarfs to the
optical  depth  depends  sensitively  on  the adopted  IMF.   For  the
standard IMF we derive a contribution of roughly 10\%, no matter what
the  magnitude  cut is,  whereas   for  the  biased  IMFs  we  obtain
contributions which  are, typically, of  23\% for the CSM1  case, 35\%
for the CSM2  case, and 22\% for the AL  simulation.  These values are
relatively constant  for large enough  magnitude cuts and,  hence, for
realistic values of the magnitude  cut they can be considered as safe.
Conversely, the  microlensing optical depth  is a robust  indicator of
the density of the microlenses.   It is important to realize that {\sl
none}  of the  adopted IMFs  is able  to reproduce  satisfactorily the
value  found  by  the  MACHO  team,  in  spite  of  the  very  extreme
assumptions  adopted for  deriving the  log-normal biased  IMFs, which
were especially taylored to reproduce the microlensing results.

Nevertheless the information that can  be derived from our Monte Carlo
simulations is  far more  complete.  A summary  of the results  can be
found in table 2, where we show the number of microlensing events, the
average  mass  of  the  microlenses,  the average  proper  motion  and
distance,  the average  tangential  velocity of  the microlenses,  the
corresponding Einstein  crossing times and,  finally, the contribution
to the optical  depth, all of them for  three selected magnitude cuts.
As it can be seen in the table 2 none of the IMFs is able to reproduce
the observed number of microlensing events (13 to 17, depending on the
selection criteria) found by Alcock  et al.  (2000).  Even in the case
of  the  CSM2  simulation,  which  corresponds  to  the  most  extreme
assumption on the  IMF, we obtain only $7\pm3$  microlensing events in
the  best of the  cases.  This  explains why  the contribution  to the
microlensing optical depth of the white dwarf population is within the
range 10\%  to 35\%, depending on  the adopted IMF,  suggesting that a
sizeable  fraction   of  the  microlensing  events   could  be  either
self-lensing in  the LMC  (Salati et al.   1999) or due  to background
objects (Green \& Jedamzik 2002).

\begin{table*}
\centering
\begin{tabular}{lcccccc}  
\hline
\hline   
\multicolumn{1}{c}{   }  &   
\multicolumn{3}{c}{Standard}   &
\multicolumn{3}{c}{CSM1} \\ 
\hline 
Age & 12 Gyr & 14 Gyr & 16 Gyr & 12 Gyr & 14 Gyr & 16 Gyr \\ 
\cline{2-4}  
\cline{5-7} 
$\langle N_{\rm WD}\rangle$  & 
$0\pm 1$ & $0\pm 1$ & $0\pm 1$ & $1\pm 3$ & $2\pm 4$ & $2\pm 4$ \\ 
$ \langle m \rangle$ $(M/M_\odot)$ 
& 0.600 & 0.565 & 0.564 & 0.588 & 0.590 & 0.586 \\ 
$\langle  \mu\rangle$  $(''\,{\rm  yr}^{-1})$ 
& 0.029 & 0.016 & 0.022 & 0.032 & 0.030 & 0.028 \\ 
$\langle d\rangle$ (kpc) 
& 1.19 & 3.09 & 2.55 & 1.59 & 1.53 & 1.78  \\  
$\langle  V_{\rm  tan}  \rangle$ $({\rm  km\,s}^{-1})$  
& 260 & 235 & 274 & 243 & 222 & 238  \\  
$\langle\hat{t}_{\rm  E}\rangle  $ (d) 
& 55.3 & 61.4 & 44.3 & 37.9 & 36.3 & 33.5 \\ 
$\langle \tau/\tau_0 \rangle$ 
& 0.110 & 0.101 & 0.090 & 0.178 & 0.209 & 0.196 \\ 
\hline 
& & & & & & \\ 
\end{tabular}  
\begin{tabular}{lcccccc}
\hline  
\multicolumn{1}{c}{  }  &   
\multicolumn{3}{c}{CSM2}   &
\multicolumn{3}{c}{AL}  \\
\hline 
Age & 12 Gyr & 14 Gyr & 16 Gyr & 12 Gyr & 14 Gyr & 16 Gyr \\ 
\cline{2-4}  
\cline{5-7} 
$\langle N_{\rm WD}\rangle$  
& $8\pm 3$ & $7\pm 3$ & $7\pm 3$ & $5\pm 5$ & $5\pm 5$ & $3\pm 3$ \\ 
$ \langle m \rangle$ $(M/M_\odot)$ 
& 0.620 & 0.598 & 0.589 & 0.642 & 0.622 & 0.650 \\ 
$\langle \mu\rangle$  $(''\,{\rm  yr}^{-1})$ 
& 0.039 & 0.093 & 0.041 & 0.043 & 0.035 & 0.026 \\ 
$\langle d\rangle$ (kpc) 
& 1.24 & 0.51 & 1.04 & 1.07 & 1.63 & 1.66  \\  
$\langle  V_{\rm  tan}  \rangle$ $({\rm  km\,s}^{-1})$  
& 230 & 224 & 202 & 219 & 272 & 203  \\  
$\langle\hat{t}_{\rm  E}\rangle  $ (d) 
& 27.0 & 21.7 & 20.9 & 38.1 & 28.8 & 48.4 \\ 
$\langle \tau/\tau_0 \rangle$ 
& 0.491 & 0.466 & 0.415 & 0.323 & 0.308 & 0.310 \\ 
\hline 
\hline
\end{tabular}
\caption{Summary results  of the  microlensing events towards  the LMC
	for  a limiting  magnitude of  $m_{\rm V}=17.5^{\rm  mag}$ and
	different halo ages and IMFs.}
\end{table*}

Another important  information which can be readily  obtained from the
simulations presented so far  is the distribution of Einstein crossing
times.   Such distributions  for the  40 independent  realizations are
shown in figure 4 for the  four cases studied here as solid lines.  We
have chosen the brighter of our magnitude cuts in order to allow for a
larger number  of events.  Note,  however, that this magnitude  cut is
exactly the  same adopted  by the MACHO  team.  Also shown,  as dashed
lines, are the distribution of Einstein crossing times obtained by the
MACHO team.  All  the distributions have been normalized  to unit area
and,  hence,  are   frequency  distributions.   The  average  Einstein
crossing times for each of the  simulations can be also found in table
2.  As it can be seen in  figure 4 the MACHO team detects microlensing
events with larger durations than those obtained in the simulations of
the log-normal IMFs  of Chabrier et al.  (1996)  and Adams \& Laughlin
(1996).  Although the  statistics is poor, we can  compare the average
crossing times  obtained here with the average  crossing time obtained
by the MACHO team which is of $76\pm 23$ days.  These average crossing
times are  respectively $\sim  61$, 36,  22 and 29  days.  In  all the
cases the typical standard deviation is of about 15 days. We emphasize
here  that these  averages  are the  result  of an  ensemble of  forty
simulations.   Clearly,  biased   mass  functions  yield  considerably
smaller average Einstein crossing times  than those of the MACHO team,
whereas a  standard IMF  yields a value  which is actually  within the
$1\sigma$  error  bars of  the  observed  one.  Since $\langle  t_{\rm
E}\rangle^2$ is an  indicator of the mass of the  lens it follows that
the  average mass  of the  lenses for  biased mass  functions  must be
different of that  of the standard one. And this  is, indeed, the case
as it  can be seen in table  2. This behavior is  not surprising since
the three log-normal mass functions partially inhibit the formation of
low mass white dwarfs.  In  all these three cases the observed average
Einstein crossing  time is beyond  the 1$\sigma$ error  bars ($\sim15$
days).   Hence, although  this  mass functions  yield a  significantly
higher number  of microlensing  events they have  considerably smaller
Einstein crossing times.

Another interesting fact which results from a careful study of table 2
has to do with the tangential velocities of the simulated microlensing
events.   As it can  be seen  the average  tangential velocity  of the
microlenses is  typical of the  halo population, with values  close to
the  canonical one  of 220~km/s.   But, on  the contrary,  the average
distances  of the  microlenses  do  depend on  the  adopted IMF.   For
instance, the microlenses produced by the standard IMF are located, on
average, at larger distances (3.14~kpc), than those of the biased IMFs
(2.12,  1.50 and  1.65~kpc, respectively).   In fact,  the  larger the
adopted  mass  cut  of the  IMF  is  (see  table  1), the  closer  the
microlenses are.  This, in  turn, translates directly into the average
proper  motions  of  the  lenses,  since, as  already  mentioned,  the
tangential  velocities  of  the  lenses  do not  differ  much  in  the
simulations reported here.  Finally,  as expected, the average mass of
the microlenses  differs considerably in  all the cases  studied here.
For the standard IMF we obtain a mass of $0.56\,M_\odot$, close to the
canonical average mass  of disk white dwarfs.  For  the biased IMFs we
obtain values of $0.58\,M_\odot$, $0.61\,M_\odot$ and $0.64\,M_\odot$,
respectively, which reflect, as  already pointed out, the adopted mass
cut for the IMF.

In  figure 5  we show  the relative  distribution (normalized  to unit
area) in magnitudes of the  whole population of white dwarfs --- solid
lines ---  and of those  white dwarfs responsible of  the microlensing
events --- dashed lines.  Also shown in each panel is the total number
of white  dwarfs found in  a typical Monte Carlo  realization.  Again,
the distributions  have been  normalized to unit  area.  As it  can be
seen in this figure, most of the white dwarfs have apparent magnitudes
brighter than $25^{\rm mag}$ and,  hence, the magnitude cut only plays
a  significant  role  in  the  case  in which  the  magnitude  cut  is
$27.5^{\rm mag}$, which is  not very realistic.  Therefore the results
quoted in table 2, do not depend much on the adopted magnitude cut.

\begin{figure*}[t]
\vspace{11cm}
\includegraphics{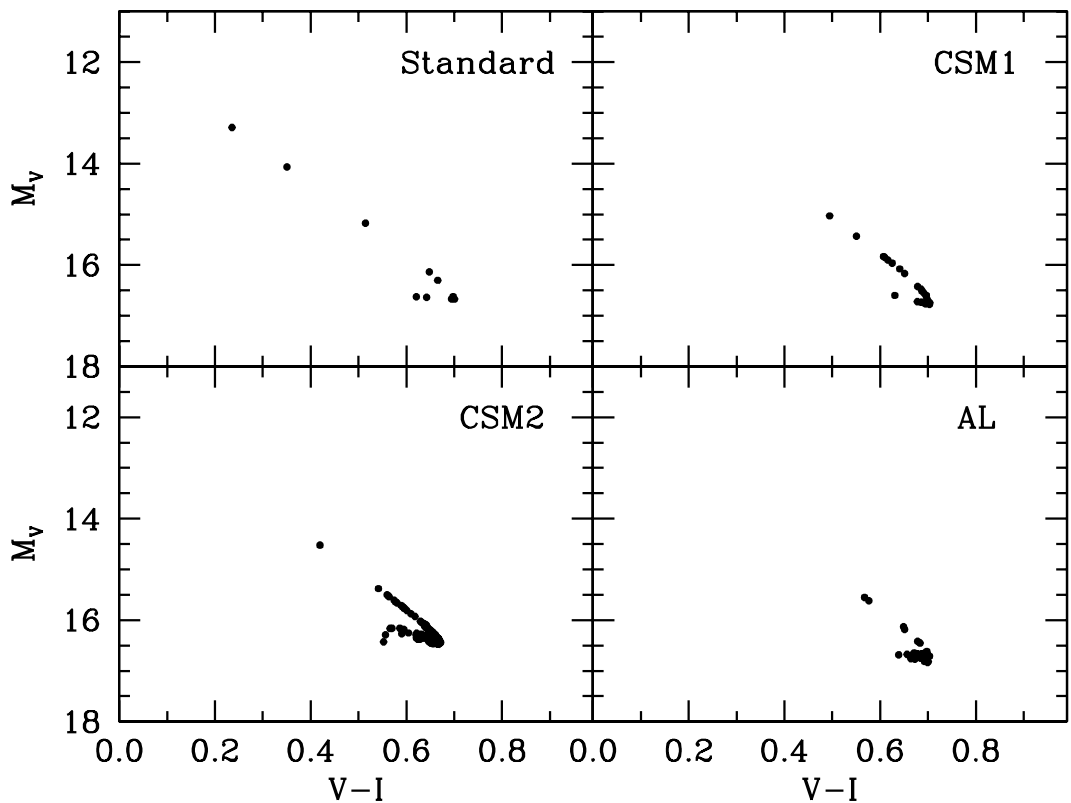}
\caption{Theoretical  color--magnitude  diagram of a typical Monte Carlo
	realization of the HDF for several IMFs.}
\end{figure*}

Finally,  we have  analyzed the  dependence  of these  results on  the
adopted age of the halo.  In order to do that we have chosen halo ages
of 12, 14,  and 16~Gyr.  In all these simulations  we have adopted the
same magnitude cut $m_{\rm  V}^{\rm cut}=17.5^{\rm mag}$.  The results
obtained in these  sets of simulations are shown in  table 3.  From an
analysis of this  table we see that the results do  not depend much on
the adopted halo age.  In particular the number of microlensing events
remains almost constant regardless of  the precise value of the age of
the  halo.  Also,  and  most  importantly, the  same  behavior can  be
observed  for  the computed  microlensing  optical  depth.  Thus,  the
actual  values  of  the  number  of microlensing  events  and  of  the
microlensing optical depth derived here are quite robust.

\subsection{The Hubble Deep Field}

The deepest optical images obtained up  to now are those of the Hubble
Deep  Field.  In  spite of  the very  small area  surveyed by  the HDF
($\simeq 4.4$~arc~min$^2$), the  limiting magnitude of $m_{\rm V}^{\rm
lim}\sim  28^{\rm mag}$ allows  to probe  a large  volume.  Therefore,
these results  complement those of the  MACHO team and  can provide us
with relevant  (and very valuable)  information about the halo  of the
Galaxy.   Nevertheless,  the  results  obtained so  far  by  different
authors are not concluding.  For instance Flynn et al.  (1996) studied
the HDF north and were  able to distinguish between stars and galaxies
down to  magnitudes as faint  as $I=26$.  Their selection  criteria in
the search of white dwarfs are summarized in the first row of table 4.
They did  not find  any object with  $V-I>1.8$ although  their results
were consistent with an upper limit of 3 white dwarfs.  Later M\'endez
et  al.  (1996)  detected  6 faint  objects  ($V \ge  25$) with  color
indexes  in the  range $-0.5<B-V<0.5$  or,  equivalently, $0<V-I<1.2$.
However,  these  objects  could  be  as well  non  resolved  galaxies.
Finally, the  most plausible analysis of  the HDF is that  of Ibata et
al.   (1999).  These  authors  determined the  proper  motions of  the
objects in the HDF from two epoch observations.  The time baseline was
2~yr.   In this  way they  were  able to  discriminate between  nearby
objects and galaxies.  The {\sl  maximum} number of white dwarfs found
by Ibata et  al.  (1999) is 4.  However,  one of these identifications
corresponds  to a  re-discovery  of a  previously  known white  dwarf,
whereas the rest of  the candidates still need spectroscopic follow-up
observations.  It  is, thus, interesting  to study which would  be the
predictions of our Monte Carlo simulator for different IMFs.

\begin{table}
\centering
\begin{tabular}{llc}
\hline
\hline
Author & Selection Criteria  & $N_{\rm Obj}$ \\
\hline
Flynn et al. (1996) &
$\left\{
\begin{array}{c}
24.63<I<26.30 \\
V-I>1.8
\end{array}
\right.$ & $<3$ \\
M\'endez et al. (1996) &
$
\left\{
\begin{array}{c}
25\leq V\leq30 \\
-0.5<B-V<0.5 \\
0<V-I<1.2
\end{array}
\right.
$ &
$\sim 6$  \\
Ibata et al. (1999) &
$\left\{
\begin{array}{c}
27<V<28 \\
-0.2<V-I<1.0
\end{array}
\right.
$
 & 4\\
\hline
\hline
\end{tabular}
\caption{Summary  of  the HDF  results.   The  authors, the  selection
	criteria and number of objects are indicated.}
\end{table}

In  figure 6  the theoretical  color--magnitude diagram  of  a typical
Monte Carlo  realization of the  HDF is shown.   As it can be  seen in
this figure most white dwarfs populate the coolest and reddest portion
of the cooling isochrone.  This is  a consequence of the fact that the
characteristic cooling times increase considerably at low luminosities
due to  both the  release of latent  heat upon crystallization  and of
gravitational energy  due to  carbon--oxygen separation (Isern  et al.
2000).   The   characteristic  ``z''-shaped  feature   at  the  lowest
luminosities is  due to the  contribution of massive white  dwarfs and
not to  the blue turn reported by  Hansen (1999) --- see  Isern et al.
(1998) and Salaris  et al.  (2000) for a  comprehensive explanation of
this feature  --- which our  cooling sequences reproduce well  but for
larger ages of  the Galactic halo.  Note as well  that for biased mass
functions  the contribution  of low  mass stars  has  been suppressed.
This, in turn, implies that these mass functions should provide a very
small number of white  dwarfs with moderately high luminosities (Isern
et al. 1998) and, indeed, this is the case.  Of course for larger mass
cuts the  number of  white dwarfs  on the hot  portion of  the cooling
isochrone should be smaller and, hence,  this is the reason why in the
AL case  and the CSM2 simulation  white dwarfs tend  to concentrate at
the hook of the cooling isochrone.

\begin{figure*}
\vspace{11cm}
\includegraphics{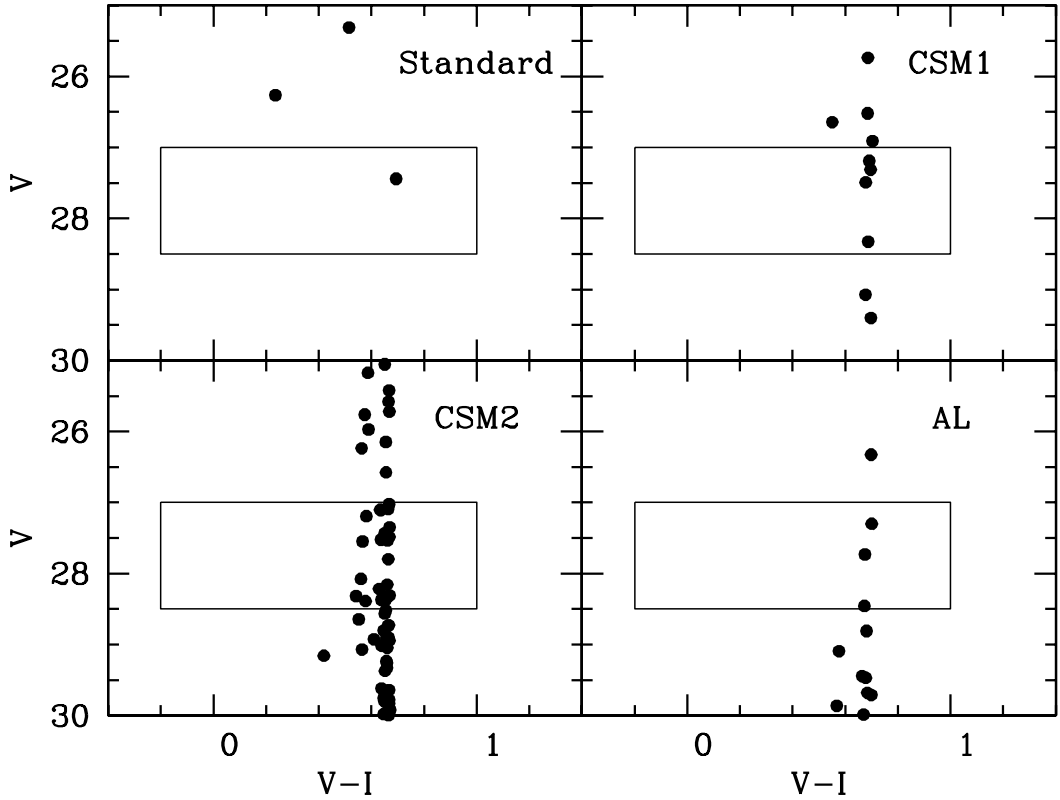}
\caption{Simulation of the HDF.   The box represents the observational
	selection criteria of Ibata et al.  (1999).}
\end{figure*}

In figure  7 a  typical realization  of the simulation  of the  HDF is
shown.  In this figure  the box represents the observational selection
criteria of  Ibata et  al.  (1999),  as shown in  table 4.   Since for
biased mass functions most white dwarfs tend to be concentrated at the
hook of the  cooling isochrone and they are  located at very different
distances  the  result  is  that  for these  cases  white  dwarfs  are
distributed along  an almost vertical  strip, which is located  at the
position of the hook, $V-I\simeq  0.6$.  This could be a potential way
to test the  IMF of the Galactic halo in addition  to the total number
of white  dwarfs found in the  field, since, as  expected, biased mass
functions provide  a considerably larger  number of white  dwarfs when
compared with the standard case.

The fraction of dark matter  of a given population is usually computed
according to:

\begin{equation}
f=\frac{3N_{\rm obj}M}{\epsilon\Omega\rho_{\rm DM}}
10^{-0.6(m_{\rm V}-M_{\rm V}+5)}
\end{equation}

\noindent where $N_{\rm obj}$ is  the number of objects in the sample,
$\rho_{\rm DM}=8.0\times 10^{-3}\, {\rm pc}^{-3}$ is the local density
of halo  dark matter, $M$ is the  average mass of the  white dwarfs in
the HDF for a given  simulation, $m_{\rm V}$ is the apparent magnitude
cut of Ibata  et al.  (1999) as  given in table 4, $M_{\rm  V}$ is the
absolute magnitude of the dimmest white dwarf in the simulated sample,
$\Omega=4.4$~arc~min$^2$  is  the  area   surveyed  by  the  HDF,  and
$\epsilon$  is  the fraction  of  white  dwarfs  within the  selection
criteria over  the total  number of white  dwarfs.  This  fraction has
been estimated to be $\epsilon\sim 0.42$ (Ibata et al.  1999), but its
real value is uncertain.  Consequently, this expression does not fully
take into account  all the biases introduced by  the selection process
and, hence, the  adopte value of $\epsilon$ can only  be regarded as a
relatively bona fide estimate.   Instead, the dark matter fraction has
been computed in the following way.

The  expected number of  objects in  the sample,  is given  by $N_{\rm
obj}=nV_{\rm eff}$, where $V_{\rm eff}$ is the effective volume of the
sample  and $n=f\rho_0/M$.   As shown  in Eq.  (11) the  usual  way to
compute   the   effective   volume    of   the   sample   is   $V_{\rm
eff}=\epsilon\frac{1}{3}\Omega  d^3$ where  $d$ is  taken to  be $\log
d=0.2(m_{\rm  V}-M_{\rm  V}+5)$.    Another,  more  accurate,  way  of
computing the effective volume is to use the $1/V_{\rm max}$ method as
it  follows.  For each  star of  the sample  we determine  the maximum
distance over which any star can contribute to the sample,

$$r_{\rm max}={\rm min}\left[\pi^{-1}(\mu/\mu_{\rm l});\,
                             \pi^{-1}10^{0.2(m_{\rm l}-m)}\right],$$  

\noindent and the minimum distance as well:

$$r_{\rm min}={\rm max}\left[\pi^{-1}(\mu/\mu_{\rm u});\,
                             \pi^{-1}10^{0.2(m_{\rm u}-m)}\right],$$  

\noindent  where  $\pi$ is  the  stellar  parallax,  $\mu$ its  proper
motion, $m$  the apparent magnitude,  $\mu_{\rm l}$ and  $\mu_{\rm u}$
are the low and high proper motion limits, if any, and $m_{\rm l}$ and
$m_{\rm u}$ are the corresponding magnitude limit.  The maximum volume
in which a star can contribute is then

$$V_{\rm max}=\frac{\Omega}{3}(r_{\rm max}^3-r_{\rm min}^3)$$  

\noindent and the number density of white dwarfs will be

\begin{table*}
\centering
\begin{tabular}{lcccccc}
\hline
\hline
IMF & $N_{\rm Obj}$ & $V_{\rm eff}$ (pc$^3$) & $\langle V/V_{\rm max}
\rangle $ & $\epsilon$ & $d$ (kpc) & $f$ \\
\hline
Standard &  $2\pm 2$ &  6360 & 0.396 & 0.69 & 3.7 & 0.04\\
CSM1     &  $4\pm 4$ &  2051 & 0.466 & 0.79 & 2.5 & 0.20\\
CSM2     & $30\pm 6$ &   874 & 0.475 & 0.68 & 1.9 & 4.16\\
AL       &  $3\pm 3$ &  2180 & 0.531 & 0.80 & 2.6 & 0.14\\
\hline
\hline
\end{tabular}
\caption{Summary of  the results for  the HDF obtained with  our Monte
	Carlo simulator for the different proposed IMFs.}
\end{table*}

$$n=\sum_{i=1}^{N_{\rm obj}}\frac{1}{V_{{\rm max}_i}}$$.  

\noindent The effective volume of the sample can be then computed as:

$$V_{\rm eff}=\frac{N_{\rm obj}}{n}$$

The  value of  $\langle V/V_{\rm  max} \rangle$  is a  measure  of the
completeness of the sample, which  is related to $\epsilon$ by

$$\epsilon=1-2\Bigg |\Big\langle \frac{V}{V_{\rm max}}
\Big\rangle -0.5\Bigg |$$

\noindent We  remind here that  for a complete and  homogeneous sample
$\langle V/V_{\rm max} \rangle=0.5$.

The results  for the HDF simulations  are also shown in  table 5.  For
each of the  cases we show the number of expected  white dwarfs in the
HDF,  $N_{\rm obj}$,  with its  corresponding standard  deviation, the
effective   volume   surveyed,   the   value  of   $\langle   V/V_{\rm
max}\rangle$, the estimated completeness,  the average distance of the
white  dwarfs in  the field,  and the  corresponding fraction  of dark
matter derived from  the simulations.  An analysis of  table 5 reveals
that the  expected number  of objects  is roughly 3  in all  the cases
except for the case CSM2, for which the value of expected white dwarfs
is much  larger.  It is interesting  to compare these  values with the
expected number of  objects if all the dark matter  of the Galaxy were
in the  form of  white dwarfs.  This  number ranges, depending  on the
selection criteria  in colors and  magnitudes, from 9 to  12.  Clearly
even biased mass functions such as  the AL and CSM1 mass functions are
not able to  fill the halo with white dwarfs.   Moreover, for the CSM2
case even if we  adopt the lower limit of 24 objects  in the field the
result is  clearly much larger than  that needed to fill  all the halo
with white dwarfs.   

On the other hand, the effective  volume surveyed by the HDF turns out
to be dependent of the adopted  model IMF.  This would not be the case
if Eq. (11)  would have been adopted since in  this case the effective
volume surveyed only depends on the absolute magnitude of the faintest
white dwarf in  all the simulated samples which turns  out to be $\sim
16.8^{\rm mag}$ --- in accordance  with the value adopted by Richer et
al.  (2000) --- and on the adopted magnitude cut ($m_{\rm V}=28.5^{\rm
mag}$).  It follows then that if  this was the case, the radius of the
effective volume would be in  all cases $\simeq 2.2$~kpc.  The average
proper  motions obtained  are  in  all four  cases  very difficult  to
measure:  $\sim   26$~mas~yr$^{-1}$,  for  the   standard  IMF,  $\sim
30$~mas~yr$^{-1}$ for  the CSM1 case, $\sim  34$~mas~yr$^{-1}$ for the
CSM2 mass function  and $\sim 32$~mas~yr$^{-1}$ for the  AL case.  The
typical standard deviation is $\pm 15$~mas~yr$^{-1}$.  It is important
to realize that these values  are in agreement with the observed value
of $24\pm 10$~mas~yr$^{-1}$ (Ibata et al.  1999).  The completeness of
the sample  is roughly  the same  in all four  cases, whereas  for the
average distance of the microlenses we have significant variations. In
particular for  the standard IMF the average  distance is considerably
larger than  in the rest  of the simulations,  being the case  CSM2 an
extreme case.   Finally, the fraction  of baryonic dark matter  in the
form of  white dwarfs for the  CSM2 case is  totally incompatible with
the observations,  whereas in the  rest of the cases  is significantly
smaller.  For  instance, for the case  the CSM1 and  AL mass functions
the fractions are comparable, and for the standard IMF the fraction of
baryonic  dark matter only  amounts to  a modest  4\%.  Finally  it is
worth  noticing here that  if we  had used  the selection  criteria of
Flynn et  al.  (1996) we would have  not found any white  dwarf in the
HDF, whereas if we have used those of M\'endez et al.  (1996) we would
have found $1\pm  1$ white dwarfs in HDF,  almost independently of the
adopted  IMF.  Hence,  it seems  that  a significant  fraction of  the
objects found by M\'endez et al.  (1996) should not be white dwarfs.

\subsection{The EROS results}

\begin{figure*}
\vspace{11cm}
\includegraphics{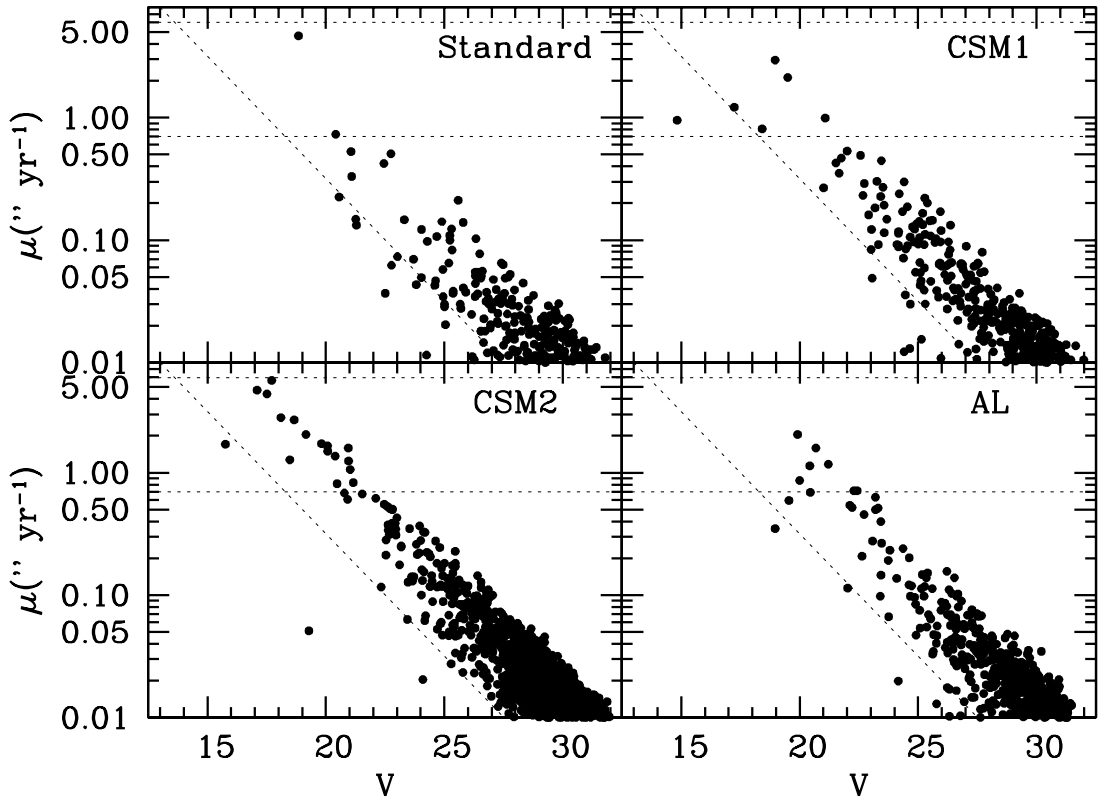}
\caption{Simulation  of the  EROS results.   See text  for  a detailed
	description.}
\end{figure*}

\begin{table*}
\centering
\vspace{0.3cm}
\begin{tabular}{lcccccc}
\hline
\hline
IMF & $N_{\rm obj}$ & $ V_{\rm eff}$ (pc$^3$) & $\langle V/V_{\rm max}
\rangle $ & $\epsilon$ & $d$ (pc) & $f$ \\
\hline
Standard &  $2\pm 1$ & 46577 & 0.210 & 0.42 & 122 & 0.01\\
CSM1     &  $6\pm 2$ & 13256 & 0.115 & 0.32 &  81 & 0.16\\
CSM2     & $12\pm 4$ &  6042 & 0.150 & 0.30 &  62 & 0.55\\
AL       &  $6\pm 4$ & 10238 & 0.140 & 0.28 &  74 & 0.17\\
\hline
\hline
\end{tabular}
\caption{Summary of the results  for the EROS experiment obtained with
	our Monte Carlo simulator for the different proposed IMFs.}
\end{table*}

The EROS  team has reported very  recently (Goldman et  al.  2002) the
results of a  proper motion survey which aimed  to discover faint halo
white dwarfs with  high proper motions.  However, they  did not detect
any candidate halo  white dwarf even if the  survey was sensitive down
to  $V\approx   21$  and   to  proper  motions   as  large   as  $\mu=
6.0^{\prime\prime} \,  {\rm yr}^{-1}$.  Moreover, they  found that the
halo white dwarf contribution cannot exceed 5\% at the 95\% confidence
level for objects with color index $1.0 \leq V-I \leq 1.5$.  It should
be noted  however that  this last result  is dependent on  the adopted
model of the Galaxy.  In particular Goldman et al.  (2002) adopted the
biased  mass  function   CSM1  and  in  order  to   compare  with  the
observational results they only simulated white dwarfs with magnitudes
in the  range $16.5 \leq M_{\rm  V} \leq 18.0$ and  color index within
$-0.5 \leq  V-I \leq  1.5$.  It is  therefore necessary to  extend the
study  to other mass  functions and,  moreover, to  the full  range of
white dwarf magnitudes and colors.  Additionally, since the results of
the EROS team  are closely connected with those of  the HDF studied in
\S 3.4  before, it is interesting to study  the results obtained
with our Monte Carlo simulator.

In figure 8  a typical Monte Carlo realization of  the EROS results is
shown, whereas  in table 6 the  average values for the  ensemble of 40
independent realizations are  also shown.  The entries in  table 6 are
the same  of table 5,  except for the  distance which is  expressed in
pc.  The selection  criteria of  the EROS  collaboration are  shown as
dashed  lines in  figure 8.  First, halo  white dwarfs  should  have a
reduced proper motion $H_{\rm V}=V+5\log \pi+5>22.5$, this restriction
is shown  in figure  8 as a  dashed diagonal line.   Additionally they
required $6.0^{\prime\prime} \, {\rm yr}^{-1} > \mu>0.7^{\prime\prime}
\, {\rm  yr}^{-1}$.  Both limits  are shown as  well in figure 8  as a
horizontal dashed lines.   As it can be seen in table  6 and in figure
8, in the region where  the EROS experiment conducted their search for
halo  white dwarfs very  few of  them would  be eventually  found.  In
particular for the case in which a standard mass function is used only
2 of them (in  the best of the cases) would be  found. This is not the
case  for the  biased mass  functions CSM1  and AL  for which  6 white
dwarfs  could be  presumably found,  whereas  for the  CSM2 case  this
number increases to 12. Since  the EROS experiment found none the only
mass function that seems to fit the observational data is the standard
one.  In  summary, both  the  results  of the  HDF,  and  of the  EROS
experiment point towards the same conclusion: that the IMF adopted for
the halo should be the standard one.

\subsection{What if...?}

\begin{figure}
\vspace{6.5cm} 
\includegraphics{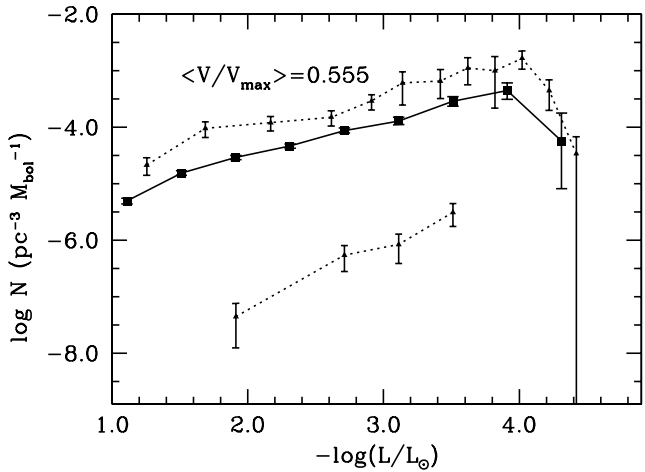}
\caption{The  halo  white  dwarf  luminosity  function  obtained  when
	assuming that  all the reported microlensing  optical depth is
	due to lenses  in the form of white dwarfs  --- solid line ---
	compared to  the observation luminosity functions  of the disk
	--- top  dotted line  --- and  of the  halo ---  bottom dotted
	line.}
\end{figure}

As it can be seen from tables  2 and 3 our results are compatible with
a halo where roughly  10\% of the dark matter is in  the form of white
dwarfs.  It is thus interesting to ask ourselves what would eventually
be the observed luminosity function  of halo white dwarfs if {\sl all}
the microlensing events (and the  associated optical depth) are due to
halo  white dwarfs.   To this  regard we  have conducted  a  series of
simulations in which  we have produced as many  white dwarfs as needed
in order to reproduce the optical depth found by the MACHO team, which
roughly corresponds to $13\pm1$ microlenses.  In all these simulations
the standard IMF  was adopted, since the biased  mass functions AL and
CSM1 studied in the previous sections produce roughly the same results
with  regard to  the white  dwarf  luminosity function.   It is  worth
noticing  as  well  that   the  ad-hoc  mass  functions  studied  here
overproduce white  dwarfs either in the  HDF or in the  EROS survey in
sharp  contrast with  the  observations.  The  white dwarf  luminosity
function  was then  computed  using the  restrictions  for the  visual
magnitude and  proper motion detailed  in \S 2.1.   As was done  in \S
3.1, forty  independent realizations of  this sample were  computed in
order to obtain the average white dwarf luminosity function along with
its  corresponding standard  deviation for  each luminosity  bin.  The
result is  shown in figure  9 as  a solid line.   For the sake  of the
comparison  the observational  disk  and halo  white dwarf  luminosity
functions  of Bergeron,  Legget \&  Ruiz (1998)  for the  disk  and of
Torres et al.  (1998) for the halo are also shown as dotted lines.  As
it can be  seen in this figure, should white  dwarfs be responsible of
{\sl  all} the  microlensing events  the halo  white  dwarf luminosity
function would be much closer to  that of the disk rather than to that
of the halo. It is, however,  worth mentioning at this point that this
is indeed an  overestimate since for at least 4  lensing events it has
been already shown  that the lenses are not in  the Galactic halo.  In
particular, one of the lenses  resides in the galactic disk (Alcock et
al.  2001a).   Moreover, three of  these are binary events  (Afonso et
al. 1988;  Bennet et al.   1996; Alcock et  al.  2001b) in the  LMC or
SMC.  It is  obvious, that it is not possible  that all lensing events
with lenses in the LMC and  SMC are binary and, hence, it follows that
a substantial fraction of single events must also be in the LMC or the
SMC as first pointed out by  Sahu (1994).  However, it is true as well
that  this   simulation  reinforces  the  result  that   most  of  the
microlensing  events reported  by the  MACHO team  are not  halo white
dwarfs and that another explanation must be found for these results.

\subsection{The role of the density profile}

In order  to check the  sensitivity of our  results with respect  to a
different density profile  we have done a final  series of simulations
in which  the density profile of  Navarro, Frenck \&  White (1997) was
adopted.  This  density  profile  is characterized  by  the  following
expression

\begin{equation}
\rho(r)=\frac{\rho_{\rm crit}\delta_{\rm c}}{(r/r_{\rm s})(1+r/r_{\rm s})^2},
\end{equation}

\noindent where $r_{\rm  s}$ is the scale radius,  $\delta_{\rm c}$ is
an adimensional characteristic density, and $\rho_{\rm crit}=3H^2/8\pi
G$ is the critical density of an Einstein--De Sitter universe.  On the
other hand,  $r_{\rm s}$ is usually defined  through the concentration
parameter $c=R_{200}/r_{\rm  s}$ as a  function of the  virial radius,
$R_{200}$.  In order to correctly  reproduce the rotation curve of our
Galaxy  the following set  of parameters  is usually  adopted: $c=10$,
$R_{200}=180$  kpc,  which results  in  $r_{\rm  s}=18$  kpc. In  this
simulation,  and as  discussed  previously, the  adopted initial  mass
function  was  the  standard  one.  We only  computed  the  number  of
microlenses towards the  LMC and not the results of the  HDF or of the
EROS experiment.

\begin{table}
\centering
\begin{tabular}{lccc}
\hline
\hline
\multicolumn{1}{c}{ } & \multicolumn{3}{c}{Standard}  \\
\hline
Magnitude & $17.5^{\rm mag}$ & $22.5^{\rm mag}$& $27.5^{\rm mag}$ \\
\cline{2-4} 
$\langle N_{\rm WD} \rangle$ & $0\pm 1$ & $0\pm 1$ & $0\pm 1$  \\
$ \langle m \rangle$ $(M/M_{\odot})$ & 0.685 & 0.660 & 0.651 \\
$\langle \mu\rangle$ $(''\,{\rm yr}^{-1})$ &
0.158 & 0.062 & 0.019  \\
$\langle d\rangle$ (kpc) &
0.33 & 1.21 & 4.25  \\
$\langle V_{\rm tan} \rangle$ $({\rm km\,s}^{-1})$ &
248 & 360 &  386 \\
$\langle \hat{t}_{\rm E}\rangle $ (d) & 
18.5 & 23.2 & 40.5 \\
$\langle \tau/\tau_0 \rangle$ &
0.064 & 0.061 & 0.060\\
\hline
\hline
\end{tabular}
\caption{Summary of  the results obtained for  the microlensing events
        towards the  LMC when  the density profile  of Navarro  et al.
	(1997) is adopted. See text for more details.}
\end{table}

\begin{figure}
\vspace{11cm} 
\includegraphics{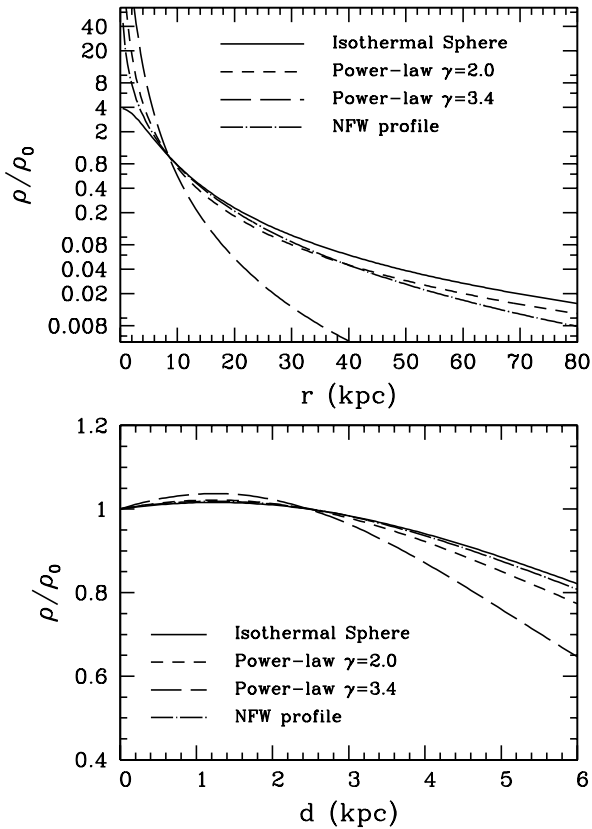}
\caption{The density  profiles used in the simulations.  The top panel
        shows  the density  profiles from  the center  of  the Galaxy,
        whereas the  bottom panel  shows the density  profiles towards
        the direction of the LMC.}
\end{figure}

The  different density  profiles  used here  are  displayed in  figure
10. The  solid line  corresponds  to the  classical isothermal  sphere
density profile,  the short dashed line  to a power  law with exponent
$\gamma=2.0$, the  long dashed line  to a power law  with $\gamma=3.4$
and  the dashed--dotted  line to  the  Navarro et  al. (1997)  density
profile. The top panel shows the  density profile as a function of the
galactocentric  $r$ coordinate,  whereas  the bottom  panel shows  the
density profile projected towards the  direction of the LMC. As it can
be  seen in  the top  panel of  figure 10,  all the  density profiles,
including that  of Navarro et  al. (1997), are much  more concentrated
towards  the  center  of   the  Galaxy  than  the  isothermal  sphere.
Additionally, the  power law  with $\gamma=3.4$ decreases  much faster
than the  other three density profiles considered  here. However, when
projected towards  the LMC, and  for the distances of  interest (those
relevant for the study of the microlenses towards the LMC, obtained in
the previous sections, 3 kpc), all four density profiles yield similar
densities.  Consequently, we do  not expect large departures from what
has been found in previous sections,  and this is indeed the case.  In
particular,  the results  obtained  using the  Navarro  et al.  (1997)
density profile are shown in table 7, and it is evident from a cursory
inspection that  the results  do not differ  much from those  shown in
table  3.   Hence,  the   same  conclusions  apply,  namely  that  the
contribution of  halo white dwarfs to the  microlensing events towards
the LMC must be small, independently of the adopted density profile.

\subsection{Miscellanea}

In order  to check the robustness  of our results we  have conducted a
series of numerical experiments in which, firstly, the distance to the
LMC was varied by a 10\% and, secondly, the tangential velocity of the
resulting white  dwarf population was also  varied by a  10\%.  In all
the cases  a standard  IMF was  used. The results  are shown  in table
8. Our  fiducial  model is  shown  in  column  3.  The  second  column
corresponds to  a model  in which  a distance of  45 kpc  was adopted,
whereas the model in  which a distance of 55 kpc was  used is shown in
column 4.  Column  5 and 6 correspond, respectively,  to the models in
which the  tangential velocities were decreased  ($V_{\rm tan}^-$) and
increased ($V_{\rm  tan}^+$) artificially  by a 10\%.  As can  be seen
there,  the differences  in all  the cases  are relatively  small. For
instance decreasing  the distance to the  LMC from 50  to 45 decreases
the contribution of white  dwarfs to the observed microlensing optical
depth by  6\%, whereas  the model in  which the contribution  of white
dwarfs to the  microlensing optical depth would be  largest is that in
which  the tangential  velocities of  all the  white dwarfs  have been
increased  by  10\%.  In  this  case  the  contribution amounts  to  a
13\%. Given all  these considerations it seems unlikely  that we could
have underestimated  the contribution of white dwarfs  to the observed
microlensing optical depth.

\begin{table}[t]
\centering
\begin{tabular}{lccccc}
\hline
\hline
\ & $45$ kpc & $50$ kpc & $55$ kpc & $V_{\rm tan}^-$ & $V_{\rm tan}^+$ \\
\hline
$\langle N_{\rm WD} \rangle$ & $0\pm 1$ & $0\pm 1$ & $0\pm 1$ & $0\pm 1$ &$0\pm 1$ \\
$\langle \mu\rangle$ $(''\,{\rm yr}^{-1})$  & 0.030 & 0.016 & 0.017 & 0.010 & 0.041 \\
$\langle \tau/\tau_0 \rangle$ & 0.063 & 0.101 & 0.086& 0.092 & 0.132 \\
\hline
\hline
\end{tabular}
\caption{Summary of  the results obtained for a  limiting magnitude of
	$m_{\rm  V}=17.5^{\rm mag}$  and  diferent values  of the  LMC
	distance and the tangential velocity}
\end{table}

\section{Discussion and conclusions}

Using  a  Monte Carlo  code  we  have  computed self-consistently  and
simultaneously  the theoretical  expectations of  the number  of white
dwarfs for  four different observational results,  namely the reported
microlensing events towards the  Large Magellanic Cloud (Alcock et al.
1997, 2000), the local halo white dwarf luminosity function (Torres et
al.  1998), the results of the  Hubble Deep Field (Ibata et al.  1999)
and the  results of  the EROS experiment  (Goldman et al.  2002).  Our
Monte Carlo  simulator takes into account all  the known observational
biases  and uses  a  thorough  description of  the  properties of  the
Galactic halo.  In  our calculations we have simulated  the halo white
dwarf population for  several halo ages (12, 14 and  16 Gyr).  We have
also used four different initial mass functions: a standard IMF (Scalo
1998),  the two  ad-hoc  initial  mass functions  of  Chabrier et  al.
(1996),  and the  biased initial  mass function  of Adams  \& Laughlin
(1996) --- see table 1.

We have  found that  none but  the standard IMF  is able  to reproduce
simultaneously the observations of the HDF, of the EROS experiment and
of the local halo white dwarf luminosity function.  In particular, the
most extreme initial  mass function of Chabrier et  al. (1996) largely
overproduces white  dwarfs for both the  EROS survey and  for the HDF.
More precisely, the  EROS experiment in their survey  did not find any
candidate white  dwarf whereas our  simulations show that if  this was
the initial mass function of the halo they should have found around 12
white dwarfs. The  same holds for the HDF, where a  maximum of 4 white
dwarfs should have been found,  whereas our simulations using this IMF
yield about 30 white dwarfs. Thus, this IMF is completely ruled out by
observations.  Regarding the more  conservative IMF of Chabrier et al.
(1996) our simulations predict  $4\pm 4$ white dwarfs for  the HDF and
$6\pm  2$  for  the  EROS  survey. Therefore,  although  this  IMF  is
marginally compatible with  the results of the HDF  and with the local
white  dwarf luminosity function,  it turns  out that  is incompatible
with  the results of  the EROS  survey.  Since  the Adams  \& Laughlin
(1996) initial  mass  function is  very  similar  to  the Chabrier  et
al. (1996) mass function the same conclusions apply to this IMF.

Regarding the microlensing  events towards the LMC we  have found that
white dwarfs  at most could be  responsible of $\sim 1$  event for the
case in which the standard IMF is adopted. This number increases up to
$\sim 5$  events for  the cases in  which the  ``moderate'' log-normal
mass  function of  Chabrier et  al. (1996)  and of  Adams  \& Laughlin
(1996) are  used.  The contribution  of these  events to  the observed
optical  depth $\tau_0$ is,  respectively, $\tau/\tau_0\sim  0.1$, 0.2
and 0.3. Hence, even if  ad-hoc initial mass functions are adopted the
total number of microlensing events  produced by halo white dwarfs and
the  corresponding optical  depth found  by the  MACHO team  cannot be
recovered  when the  observational  biases and  other constraints  are
taken into  account. Moreover, the  average duration of  the simulated
microlensing events for all the samples in which biased mass functions
were used is  considerably smaller than the observed  one. In summary,
the ad-hoc  initial mass  functions of Chabrier  et al. (1996)  and of
Adams \& Laughlin (1996), which  were tailored to fit the observations
of the  MACHO team,  fail to reproduce  the rest of  the observational
constraints {\sl and} do not yield the required number of microlensing
events  (and, thus,  do not  significantly contribute  to  the optical
depth).   Consequently  they  should  be disregarded  until  new  (and
unlikely) evidence could eventually appear.

We would like to emphasize at  this point that all the above mentioned
results  are  almost  completely   independent  of  the  adopted  halo
age.  Moreover, in  order  to check  whether  or not  our results  are
sensitive to  the density  profile of the  halo, we have  extended our
simulations using the  density profile of Navarro et  al.  (1997).  We
have  found that  the microlensing  results are  not sensitive  to the
choice of the density profile because at the distances of interest all
the density profiles studied here are coincident.

In summary, our  results strongly suggest that a  sizeable fraction of
the dynamical mass in the halo of  our Galaxy cannot be in the form of
old hydrogen-rich  white dwarfs.  More  specifically, for the  case in
which a standard initial mass  function is adopted this fraction turns
out to be of the order of 4\% if we trust the results of the HDF or as
low as 1\% if  the results of the EROS team are  adopted.  In any case
the percentage obtained here is well below the claim of the MACHO team
that 20\% of the dark halo of the Galaxy is tied up in half-solar mass
stellar bodies (Alcock et al.  2000), since the constraints set by red
star counts (Flynn  et al.  1996) discard the  possibility of low-mass
main sequence stars.   On the other hand our  results are in excellent
agreement not  only with the observations  of the HDF and  of the EROS
experiment but  also with  the theoretical estimates  of Graff  et al.
(1998), $\la 10\%$, Flynn et  al. (2001), $\la 2\%$, and more recently
Goldman  et al.  (2002),  $\la 4\%$,  and Zhao  (2002), $\la  4\%$. Of
course, the  real nature of  the reported microlensing  events remains
still a mistery  and, hence, the last word on this  issue has not been
told yet.


\begin{acknowledgements}
Part of  this work was  supported by the MCYT  grants AYA04094--C03-01
and 02, by the MCYT/DAAD grant HA2000--0038, and by the CIRIT.
\end{acknowledgements}


\end{document}